\documentstyle{europhys}

\def\etal{{\hbox{{\tenit\ et al.\/}\tenrm :\ }}}

\def\stars{\bigskip\centerline{***}\medskip}

\newif\ifboo \boofalse

\def\Review#1{\boofalse{\it #1},}
\def\Name#1{{\sc #1},}
\def\Vol#1{\ifboo Vol. {\bf #1}\else{\bf #1}\fi}
\def\Year#1{\ifboo #1\else(#1)\fi}
\def\Book#1{\bootrue{\it #1},}
\def\Page#1{\ifboo {\rm p. #1}\else{\rm #1}\fi}
\input epsf

\begin{document}

\euro{00}{0}{1-$\infty$}{1999}
\Date{16 July 1999}
\shorttitle{C. SIMAND \etal INHOMOGENEOUS TURBULENCE.}

\title{Inhomogeneous turbulence in the vicinity of a large scale coherent
vortex}
\author{C. Simand, F. Chill\`a, J.-F. Pinton}
%\footnote{author for correspondance: {\tt pinton@physique.ens-lyon.fr}}}
\institute{
     Laboratoire de Physique de l'\'Ecole Normale Sup\'erieure de Lyon, UMR
5672 \\
     46, all\'ee d'Italie, F-69007 Lyon, France}

\pacs{
\Pacs{47}{32.-y}{Rotational flow and vorticity}
\Pacs{47}{32.Cc}{Vortex dynamics}
\Pacs{67}{40.Vs}{Vortices and turbulence}
      }
\maketitle

%\draft
%\title{Intermittency in the vicinity of a large scale coherent vortex}
%\author{Catherine Simand, Francesca Chill\`a, Jean-Fran\c{c}ois Pinton}
%\address{Laboratoire de physique - CNRS URA 1325, Ecole Normale
%Sup\'erieure
%de Lyon,\\ 46, all\'ee d'Italie, 69364 Lyon cedex 7 - France}
%\date{Draft: \today}
%\maketitle

\begin{abstract}

We study the statistics of turbulent velocity fluctuations in the
neighbourhood
of a strong large scale vortex at very large Reynolds number.
At each distance from the vortex core, we observe that the
velocity spectrum has a power law ``inertial range'' of scales and that
intermittency -- defined as the variation of the probability
density function (PDF) of velocity increments as the length of the increment
is varied -- is also present. We show that the spectrum scaling exponents and
intermittency characteristics vary with the distance to the
vortex.  They are also influenced by the large scale dynamics of the vortex.

\end{abstract}

%%%%%%%%%%%%%%%%%%%%%
%INTRO
%%%%%%%%%%%%%%%%%%%%%

\section{Introduction}
Much efforts have been devoted to the study of high Reynolds number
turbulence, assuming the properties of local homogeneity and isotropy.
Under these assumptions it has been shown~\cite{mon71} that, in
between the integral scale at which energy is fed into the flow and
the dissipative scale at which viscosity smoothes out the velocity
gradients, there exists an ``inertial range'' where: {\it (i)} the
velocity spatial power spectrum follows a power law $E(k) \sim \epsilon^{2/3}
k^{-5/3}$, {\it (ii)} the energy transfer rate $\epsilon$ is related to
the third order structure function
$S_{3}(r) = \langle \delta u^{3}(r) \rangle = - \frac{4}{5} \epsilon  r$.
The first point has been verified experimentally in numerous
experiments and constitutes the main success of
Kolmogorov's K41 mean field theory~\cite{fri95}. The second point is
generally admitted although experimental verifications are
rarer~\cite{gagne}. In addition, it has been observed  that the
probability
distribution of the amplitudes of velocity increments depends on the
width
of the increment. It varies from a gaussian PDF at integral
scale to the development of stretched exponential tails at the
smallest scales. This feature  defines intermittency, and traces
back to the non-uniformity in space of the energy transfer
rate. These three observations
% ``$-5/3$''
%velocity spectrum, Kolmogorov's four-fifth law for the third order
%structure function and existence of intermittency
are the building
blocks of the studies of homogeneous, isotropic tubulence (HIT).

It is the aim of our work to study experimentally how these
observations are modified in a situation of inhomogeneous turbulence.
Partial studies have been made near weak vortices~\cite{Chilla}, in
inhomogeneous wakes~\cite{Wesfreid,Camussi} or near a boundary
layer~\cite{Leveque,Toschi}.
Here we study turbulence in the vicinity of a very intense large scale vortex.
Our motivations are twofold: first, most high Reynolds numbers flows are
inhomogenous and the question of the ``universality'' of the features of
HIT should be addressed. A second motivation is that it has been proposed
that such coherent structures play a significant role in the statistical
properties of turbulence~\cite{tsinober,andreotti,jimenez,malecot}.

%%%%%%%%%%%%%%%%%%%%%
%SETUP
%%%%%%%%%%%%%%%%%%%%%

\section{Experimental set-up and measurements}
The experimental apparatus and measurement techniques are those described
in~\cite{pin98}. A strong axial vortex is formed in the gap between
two corotating disks -- fig.~1. The control parameters of the flow are the disks angular velocities: $\Omega_{1}$, $\Omega_{2}$,
and the mean integral Renolds number:
$Re =  {R^{2} \sqrt{\Omega_{1}^{2} + \Omega_{2}^{2}} }/{\nu}.$
Essentially two regimes are observed. The first one, hereafter
labeled (GR), is when $\Omega_{1} \sim \Omega_{2}$: the core of the
flow has a strong global rotation and a stable large scale axial vortex
is present. On the other hand, when one disk rotates at a much faster
rate than the other, the flow is dominated by differential
rotation; a strong axial flow is produced and one observes
intermittent sequences of formation and bursting of large scale axial
vortices~\cite{pin98} -- in the following, we label (DR) this regime.
 In each case, the integral Reynolds number of the flow exceeds
$5 \, . 10^{5}$ so that the flow is `turbulent'.

\begin{figure}
    \begin{center}
        \epsfysize=3cm
        \epsfbox{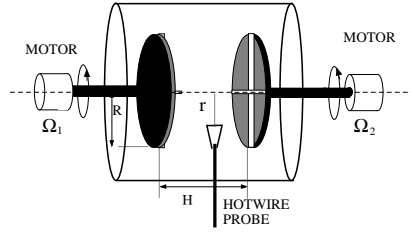}
    \end{center}
\caption{Experimental setup. The fluid is confined in a cylindrical
vessel of height 40~cm and radius 11.7~cm. The disks, of radius
$R=9.8$~cm are set a distance $H=30$~cm apart.
They are driven by two d.c. motors whose rotation rates $\Omega_1,
\Omega_2$ are adjustable in the range $[15, 45]$~Hz.
The hot-film probe is
located just above the midplane, at variable distance $r$ from the
axis of rotation. }
\label{fig1}
\end{figure}

Local velocity measurements are performed as in \cite{pin98}.
%in air using a TSI IFA100 constant
%temperature anemometer and a TSI 1260A-10 subminiature hot-film
%probe with a sensing element $25\,\mu$m thick and $0.5$ mm long.
The hot-film probe
is set parallel to the axis of rotation, so that it measures
$u_{r\theta} = \sqrt{u_{r}^{2} + u_{\theta}^{2}}$ (here, $(r,\theta,
z)$ are the conventional cylindrical coordinates with $z$ parallel to
the rotation axis). The sampling frequency $Fs$ is 78125~Hz for measurements in
the (DR) regime and $39062.5$~Hz in (GR) -- In each case, at least $10^7$ data
points are collected. We stress that although the flow has
in most  locations a well defined mean value $\overline{u}$ of the velocity
$u_{r\theta}$ and fluctuation levels $u_{\rm rms}$ comparable or less to
thoses reported in jet turbulence, we do not attempt to use and
justify a `Taylor hypothesis' here and we analyse our result as time
series.

%%%%%%%%%%%%%%%%%%%%%
%% RESULT 1 : spectra
%%%%%%%%%%%%%%%%%%%%%

\section{Spectra}
A first question regarding such a turbulence concerns the range of scales of
motion and the energy distribution among them. In fig.~2 we show the power
spectra of velocity time series recorded at different distances from the
rotation axis. It can be observed, as a first sign of the flow
inhomogeneity, that the spectra depend very much on the location of
the probe and on the flow configuration.

\begin{figure}[h]
\begin{center}
    \epsfysize=5cm
    \epsfbox{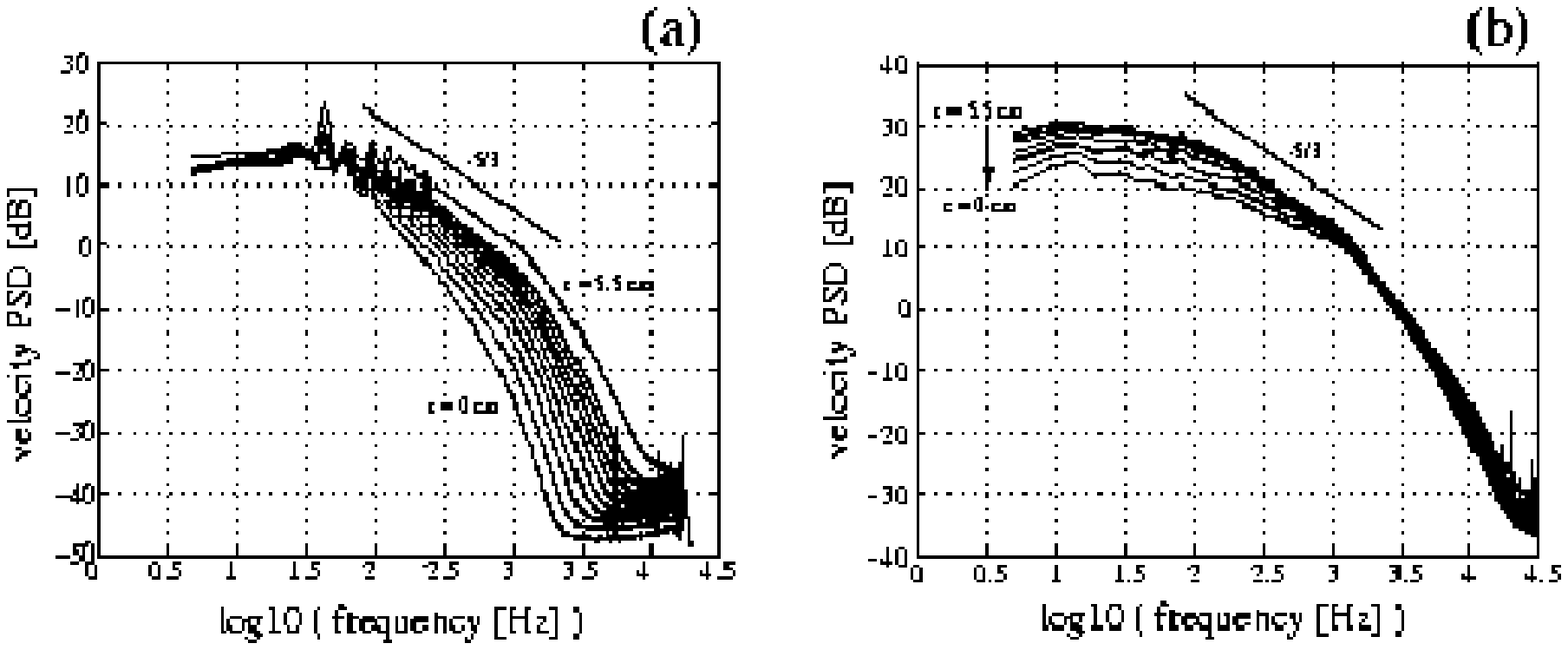}
\end{center}
    \caption{Power spectra of velocity times series recorded at distances
    $(0$, $0.5$, $1$, $1.5$, $2$, $2.5$, $3$, $3.5$,
    $4$, $4.5$, $5.5)$~cm from the rotation axis.
    (a): $\Omega_{1} = \Omega_{2} = 30$~Hz; (b) :
    $\Omega_{1} = 40$~Hz, $\Omega_{2} = 12$~Hz. }
    \label{fig2}
\end{figure}

In all cases, we observe the existence of an `inertial range' $\omega \in
[\omega_{I}, \omega_{T}]$ where the spectrum follows a power law
$E(\omega) \propto \omega^{-\alpha}$.  However both the range
$\omega_{I}/\omega_{T}$ and the exponent $\alpha$ vary. \\
- In (GR), $\alpha$ increases from about 1.65 in the outer region, to 2.50
at the
rotation axis; meanwhile the scaling domain is also reduced. A more
detailed analysis of the spectra shows that the energy in the larges scales
remains approximately constant while both the frequency and the energy at
the end of
the 'inertial range' decrease with $r$. As a result, the proximity of the
stable
vortex structure reduces the range of scales of the turbulent
motion and their energy content. This is consistent with earlier findings~\cite{mel93,hos94} which have shown that global rotation tends to
decrease the intensity of small scales motion. \\
- In (DR),
$\alpha$ decreases from again about 1.65 in the outer region, to
$1.1$ at the rotation axis. The scaling range does not vary.
 Here, one observes that the energy in the dissipative region is almost
independent of the position $r$ of measurement. It is the energy measured
in the large scales that decreases with $r$.

We thus find that the dynamics of the large scales influences the entire range
of scales. It is surprising to observe that some self-similarity is retained
since the spectra follow a power law (most curves exhibit almost a decade of
scaling behaviour): there is no scale separation where one would observe the
pathologies of the injection at large scales and where the turbulence would
recover the characteristics of HIT at small scales.

%%%%%%%%%%%%%%%%%%%%%
%% RESULT 2 : SF3
%%%%%%%%%%%%%%%%%%%%%
\section{Third order structure function}
The energy transfer across the scales is usually illustrated by
the behaviour of the
third order structure function. In HIT, this is justified by the
K\'arm\'an-Howarth relationship. But an anistropic equivalent can be
established~\cite{fri95}. We show in fig.~3 the evolution of the third order
moment of the velocity increments $S_{3}(\tau) = \langle \left( u(t) - u(t+\tau)\right)^{3} \rangle_{t}$, computed at different distances from the rotation
axis. \\
- In (GR), we observe that in the outer region, $S_{3}(\tau)$ is a
negative, bell-shaped curve. Negative values of the skewness indicate
that the cascade proceeds in the forward direction, from the large
scales to the dissipative ones, as in HIT. At
such limited Reynolds numbers, one cannot expect a plateau in plots of
$S_{3}(\tau)/\tau$ (even less with value $-4/5$), but the maximum of
$|S_{3}(\tau)/\tau|$ is consistent with the value of the power consumption per
unit mass $\epsilon$, measured from the electric consumption of the
motors~\cite{lab96}.
On the contrary going towards the axis the amplitude of $S_{3}(\tau)$
diminishes
rapidly. Near the vortex core, the skewness eventually becomes
positive in an intermediate range of increments. This indicates that
the cascade to small scales is first reduced, and then reversed by
the influence of the coherent rotation of the vortex. Energy is accumulated
in the large scale where it stabilizes the dynamics of the large scale
vortex.\\
\begin{figure}[h]
\begin{center}
    \epsfysize=4cm
    \epsfbox{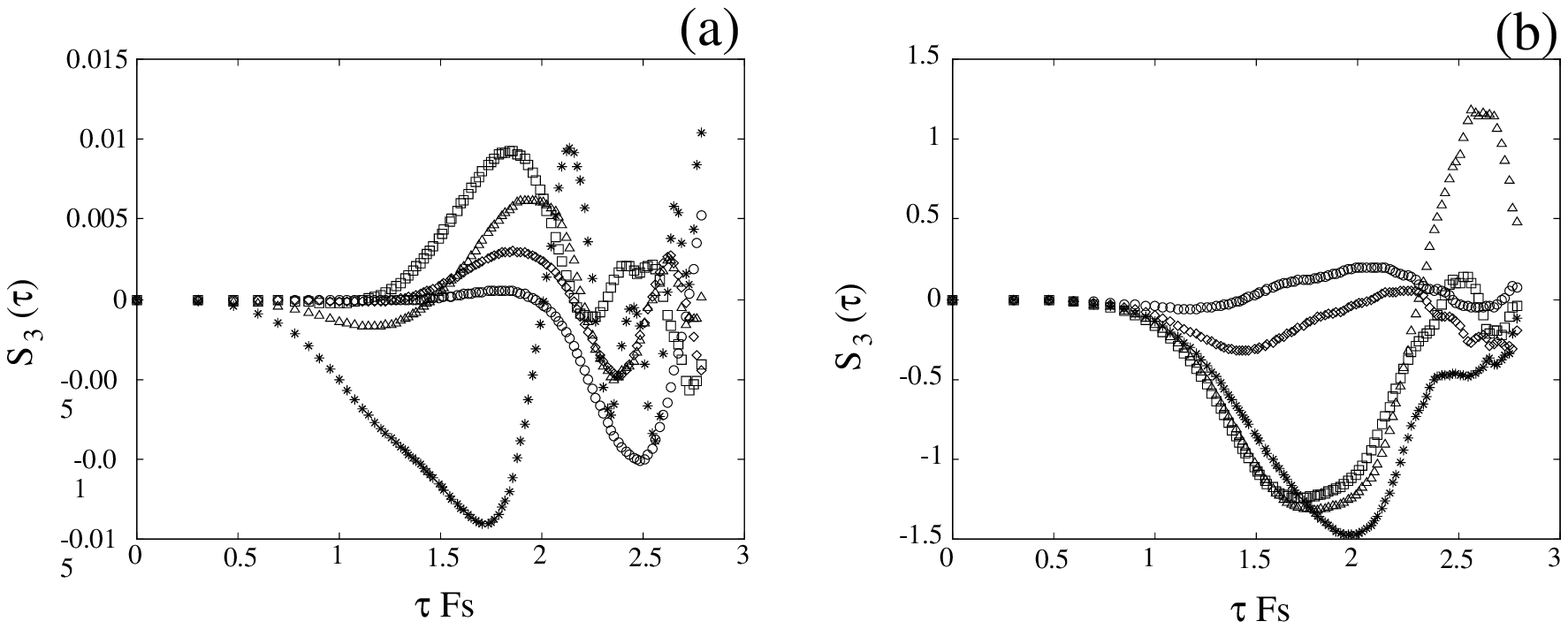}
\end{center}
\caption{Evolution with increment length of the third order moment of the velocity. (a): (GR) regime; (b): (DR) regime. $S_{3}(\tau)$ is computed at distances $r = 0.5(\circ), \; 1.5(\Diamond), \; 2.5(\Box), \; 3.5(\triangle), \; 4.5(\star)$~cm from the axis. The horizontal axis is in units of the sampling frequency.}
\label{fig3}
\end{figure}
- This picture is quite different in the irregular
(DR) regime. There, except at the rotation axis, the skewness has the
behaviour traditionnally observed in HIT. The cascade proceeds forward;
the maximum shows little variation with the location of the probe and
has a value which is, again, consistent with the motors power
consumption. At the axis, $S_{3}(\tau)$ drops suddenly; even its sign
becomes unclear and it may even be slightly positive in a range of
scales.

Altogether, the analysis of the third order moment shows that the
structure and dynamics of the flow at large scale strongly influences
the entire energy transfer across the scales.

%%%%%%%%%%%%%%%%%%%%%
%% RESULT 3 : INTERMITTENCY
%%%%%%%%%%%%%%%%%%%%%

\section{Intermittency}
A first observation is that intermittency is always present,
regardless of the flow regime and probe location. Indeed, for every
measurement, the PDFs of
velocity increment change from a quasi-gaussian shape at integral scale
to the development of stretched exponential tails at small scales,
see fig.4 (the PDFs shown in this figure are for signed quantities). 

\begin{figure}[h]
\begin{center}
    \epsfysize=4cm
    \epsfbox{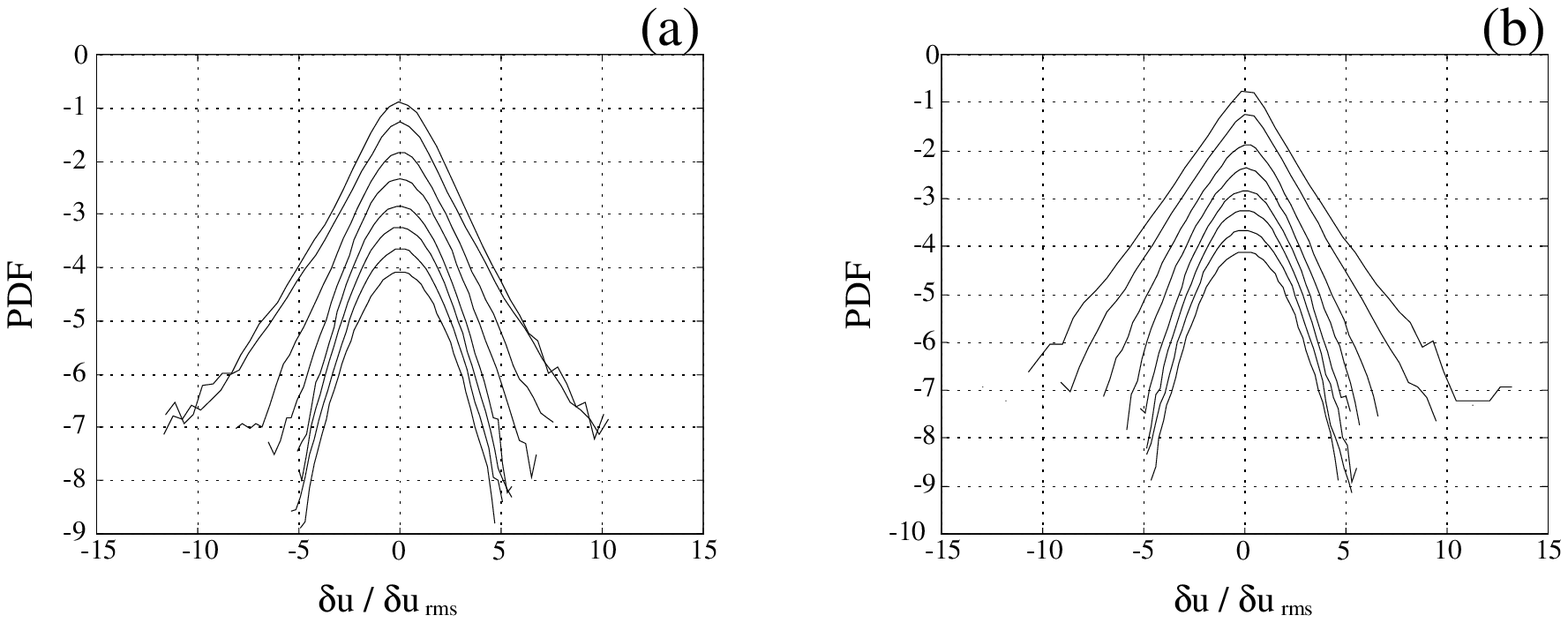}
\end{center}
\caption{Evolution with increment length of PDFs of the velocity increments.
(a): (GR) regime; (b): (DR) regime. The PDFs are computed at distance $r =
2.5$~cm
from the axis and for increments $\tau=(1,5,21,49,116,271,635,1485)/Fs$.
Values of the velociy have been normalized by the $rms$ fluctuation level.
The curves are translated for clarity.}
\label{fig4}
\end{figure}

However, as it has now become customary, we analyse here intermittency through the behaviour of
moments of the absolute values of the velocity increments:\\
$ S_{p}(\tau) = <|\delta u (\tau) |^{p} > = < |u(t+\tau) -u(t) |^{p}>_{t} $.
A feature
common to all flow configurations and every distance to the axis,
is that the evolution of $S_{p}(\tau)$ is well modeled by
a self-similar multiplicative cascade in the
sense that one can write:
$$
\log( S_{p}(\tau) ) = H(p) n(\tau) \; \; \; ,
$$
in a large interval of increments~\cite{castaing,pinton}. Here, $H(p)$ is
the Laplace
transform of the cascade propagator~\cite{casdub} which reflects the
hierarchy of
amplitudes of the velocity increments and $n(\tau)$ describes the speed at
which
the cascade proceeds along the scales.  Experimentally, we have checked the
above
relation in the  following manner: \\
{\it (i)} the plot of one moment against an other yields an extended
scaling region,
with exponent $H(p)/H(q)$ if one plots $S_p$ vs. $S_q$ -- this is the ESS
property~\cite{ben93}.
Since the decomposition of a moment $S_{p}(\tau)$ into the  two functions
$H$ and $n$
is defined up to an arbitrary multiplicative  constant, we set $H(3) = 1$.
Unlike the context of HIT, where this  choice relies on the expected
scaling of
the third order structure function, it is here {\it a priori} arbitrary.\\
{\it (ii)} once that first step is done, and the values $H(p)$ obtained,
one computes for every order the function $n_{p}(\tau) \equiv
log(S_{p}(\tau))/H(p)$.
We so verify that $n_{p}(\tau)$ is independent of $p$ and obtain a unique
function
$n(\tau)$ for every measurement.

\begin{figure}[h]
\begin{center}
    \epsfysize=4cm
    \epsfbox{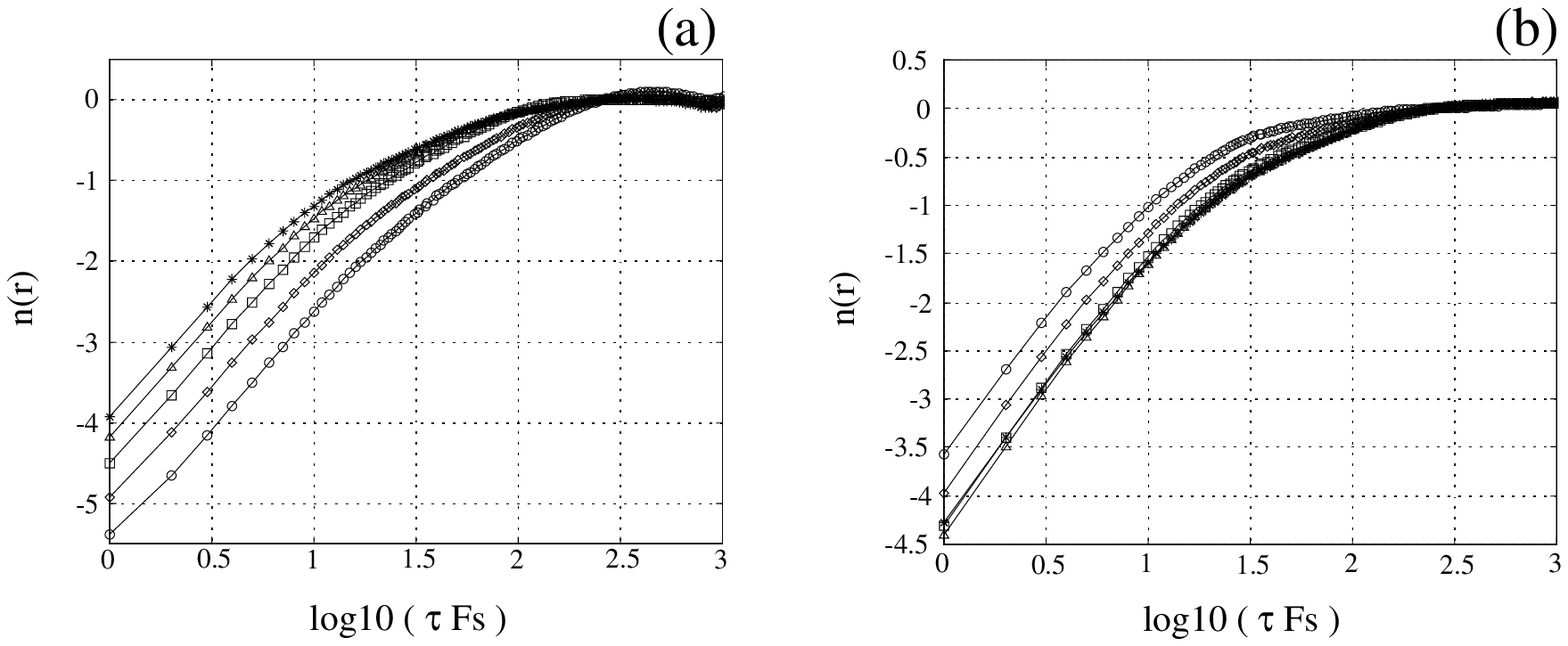}
\end{center}
    \caption{Evolution of the cascade depth. (a): (GR) regime; (b): (DR)
regime.
    $n(\tau)$ is computed at distances $r = 0.5(\circ), \; 1.5(\Diamond),
\; 2.5(\Box),
    \; 3.5(\triangle), \; 4.5(\star)$~cm from the axis.}
    \label{fig5}
\end{figure}

Following this procedure, the results are:\\
- Up to experimental errors, the functions $H(p)$ do not depend on the
measurement location nor on the flow regime. The values
($H(p)=0.42,0.70,1.27,1.51,1.73 \; {\rm for} \; p=1,2,4,5,6$) are those
reported in HIT in numerous experiments~\cite{fri95,arn95}.\\
- The functions $n(\tau)$ do depend on the measurement location and flow
regime.
They are shown in figure~5 (the curves are normalized so that $n = 0$ at
integral
time delays). We do not discuss hereafter the functional form of the
$n(\tau)$ curves but the
relative behaviour of the curves in each regime, when $r$ is varied. In (GR)
one observes that in the outer region a larger range of scales is
covered with a given number of cascade steps $n$ than in the vicinity
of the vortex core. The slopes $dn(\tau)/d\tau$ increase as $r$
decreases, meaning that increasing number of steps are needed to
cascade between any two given scales. Again, this is
consistent with the idea that rotation prevents the 3D cascade and
with our previous observation that it reduces the energy transfer to smaller
scales. An inverse effect is observed in the (DR) regime: as
measurements are performed closer to the rotation axis, $n(\tau)$
increases and is less steep, meaning that a wider range of scales is
reached in a given number of cascade steps. This is consistent with a
reduced slope of the velocity power spectra since such an efficient
cascade provides an enhanced distribution of energy in the smaller
scales.

%%%%%%%%%%%%%%%%%%%%%
%% CONCLUSION
%%%%%%%%%%%%%%%%%%%%%

\section{Concluding remarks}
Our foremost original observation is that some major
caracteristics of HIT are retained in the vicinity of large scale
coherent vortex structures: {\it (i)} the spectra develop
an ``inertial range'' in the sense that power law scaling regions
exist, {\it (ii)} the 3rd order structure functions yield consistent
descriptions of the energy transfers and {\it (iii)} intermittency is
always present and can be adequately described by a self-similar
multiplicative cascade model. Second is that the structure and dynamics
of the large scales influence the entire range of scales of motion.
The combined study of power spectra, third order structure function
and cascades parameters $H$ and $n$, give valuable informations
that show that the turbulent cascade is reduced in the presence of
global rotation and enhanced in the vicinity of very unstable vortex
structures. Note that in the (GR) regime,
eventhough the energy cascade is reversed and the spectra evolve towards
a ``$-3$'' slope in the inertial
zone, intermittency is observed: this a sharp difference with 2D
turbulence~\cite{paret} where no intermittency is detected although
deviations from Gaussianity may be present~\cite{boff99}.

%More detailed measurements are underway, in particular concerning the structure of the large scale flow, in order to refine the link between small scale characteristics and global flow configuration. This is of interest, both for the understanding of the turbulent `cascade' but also for applications such as in geophysical flows.

%%%%%%%%%%%%%%%%%%%%%
%% THANKS
%%%%%%%%%%%%%%%%%%%%%
\stars
%\vskip-12pt
It is a pleasure to acknowledge many useful discussions with B. Andreotti
and B. Castaing.

%%%%%%%%%%%%%%%%%%%%%
%%%%%%%%%%%%%%%%%%%%%

\stars
%
%%%   Bibliography environment begins here. You can use the macros \Name{},
%%%   \And, \Book{} or \Review{}, \Vol{}, \Year{} and \Page{}, to type your
%%%   references.
%

%%%%%%%%%%%%%%%%%%%%%
%%%%%%%%%%%%%%%%%%%%%

\end{document}